\documentclass[twocolumn,english,preprintnumbers,apj,floatfix]{emulateapj}

\newcommand{\beq}{\begin{equation}}
\newcommand{\eeq}{\end{equation}}
\newcommand{\bdm}{\begin{displaymath}}
\newcommand{\edm}{\end{displaymath}}
\newcommand{\bea}{\begin{eqnarray}}
\newcommand{\eea}{\end{eqnarray}}
\newcommand{\bt}{\begin{tabular}}
\newcommand{\et}{\end{tabular}}
\newcommand{\lan}{\langle}
\newcommand{\ran}{\rangle}
\newcommand{\xv}{{\bf x}}

\def\Ms{\, h^{-1} \, {\rm M}_{\odot}}
\def\Mpc{\, h^{-1} \, {\rm Mpc}}
\def\Mlim{M_{lim}} 
\def\Gpc{\, h^{-1} \, {\rm Gpc}}

\def\kMpc{\, h \, {\rm Mpc}^{-1}}

\def\fNL{f\!_{N\!L}}
\def\dfNL{\Delta f\!_{N\!L}}

\begin{document}

\preprint{FERMILAB-PUB-06-324-A}

\title{Primordial non-Gaussianity and Dark Energy Constraints from Cluster Surveys}

  \author{Emiliano Sefusatti\altaffilmark{1}}
  \author{Chris Vale\altaffilmark{1}}
  \author{Kenji Kadota\altaffilmark{1,2}}
  \author{Joshua Frieman\altaffilmark{1,3}}
  \email{emiliano@fnal.gov}
  \altaffiltext{1}{Particle Astrophysics Center, Fermi National Accelerator Laboratory, 
Batavia, IL 60510-0500}
  \altaffiltext{2}{William I. Fine Theoretical Physics Institute, University of Minnesota, Minneapolis, 
MN 55455}
  \altaffiltext{3}{Kavli Institute for Cosmological Physics and Department of Astronomy \& 
Astrophysics, The University of Chicago, Chicago, IL 60637}

\begin{abstract}

Galaxy cluster surveys will be a powerful probe of dark energy. At the same time, cluster abundance 
is sensitive to any non-Gaussianity of the primordial density field. It is therefore possible that 
non-Gaussian initial conditions might be misinterpreted as a sign of dark energy or at least degrade 
the expected constraints on dark energy parameters. To address this issue, we perform a likelihood 
analysis of an ideal cluster survey similar in size and depth to the upcoming South Pole Telescope/Dark 
Energy Survey (SPT-DES). We analyze a model in which the strength of the non-Gaussianity is parameterized 
by the constant $\fNL$; this model has been used extensively to derive Cosmic Microwave Background (CMB) 
anisotropy constraints on non-Gaussianity, allowing us to make contact with those works. We find that 
the constraining power of the cluster survey on dark energy observables is not significantly diminished 
by non-Gaussianity provided that cluster redshift information is included in the analysis.  We also find 
that even an ideal cluster survey is unlikely to improve significantly current and future CMB 
constraints on non-Gaussianity. However, when all systematics are under control, it could constitute a 
valuable cross check to CMB observations.

\end{abstract}


\maketitle

\section{Introduction}

Of the many fascinating discoveries in cosmology over the last decade, perhaps none have aroused 
more interest than the discovery of the accelerated expansion of the Universe \citep{Riess:1998cb,
Perlmutter:1998np}. Probing the nature of the dark energy thought to be driving this acceleration has 
become a top priority for the community, and among the promising tools under consideration are surveys 
of galaxy clusters. Since the number of clusters as a function of redshift and mass depends on both 
the growth of structure and on the volume of space, the cluster abundance is sensitive to the matter 
density, the density fluctuation amplitude, and the expansion history of the Universe. For this 
reason, upcoming cluster surveys will be powerful probes of cosmology \citep[e.g.][]{Haiman:2000bw,
Holder:2001db,Battye:2003bm,Molnar:2003eq,Wang:2004pk,Rapetti:2004aa,Marian:2006zp}.  

Although constraining dark energy is a leading motivator for much of the interest in cluster surveys, 
it is worth noting that the cluster abundance is potentially sensitive to various cosmological 
parameters beyond those associated with dark energy. For example, it has been recognized for some 
time that slight deviations from Gaussianity in the primordial matter distribution would cause a 
significant change in the high mass tail of the halo distribution \citep{Lucchin:1987yv,
Colafrancesco:1989px,Chiu:1997xb,Robinson:1998dx,Robinson:1999se,Koyama:1999fc,Robinson:1999wh,
Matarrese:2000iz}. In this paper, we use a maximum likelihood analysis to investigate the extent to
which dark energy constraints from cluster surveys are degraded by including the possibility of 
non-Gaussian initial conditions, in particular when considered within the limits allowed by present 
and future CMB observations.

The specific form of non-Gaussian initial conditions we consider here is of the local type, described 
in position space by a primordial curvature perturbation of the form \citep{Verde:1999ij,Komatsu:2001rj}
\beq\label{fNLmodel}
\Phi(\xv)=\phi(\xv)+\fNL[\phi^2(\xv)-\lan\phi^2(\xv)\ran]
\eeq
where $\phi(\xv)$ is a Gaussian random field and the degree of non-Gaussianity is parameterized in terms 
of the constant $\fNL$. For this model, tight constraints of the order of $\dfNL\sim 40$ are provided 
by CMB observations \citep{Komatsu:2003fd,Creminelli:2005hu,Spergel:2006hy,Chen:2006ew}, while constraints 
that are somewhat weaker but that are closer in physical scale to that of clusters are expected from 
higher-order galaxy correlations \citep{Scoccimarro:2003wn}. From a theoretical point of view, the 
non-Gaussian model of Eq.~(\ref{fNLmodel}) is motivated in part by studies of the generation of density 
perturbations in inflationary scenarios; while single-field inflation models typically predict an 
unobservably small value for $\fNL$ \citep[e.g.][]{Acquaviva:2002ud,Maldacena:2002vr}, multi-field 
inflation models can lead to much higher values \citep[e.g.][]{Lyth:2002my,Dvali:2003em,Zaldarriaga:2003my,
Creminelli:2003iq,Arkani-Hamed:2003uz,Alishahiha:2004eh,Kolb:2005ux,Sasaki:2006kq}. For a review, 
see \citet{Bartolo:2004if}.

While we believe it is worthwhile to keep an open mind to other forms of non-Gaussianity which may not 
be properly described by the simple expression in Eq.~(\ref{fNLmodel}), and which might make the 
extrapolation of current CMB constraints to cluster scales less straightforward than we assume here 
\citep[see, e.g.,][]{Mathis:2004vi}, we note that the physical scale probed by clusters differs from that of 
the Planck survey by roughly a factor of two, so that the two probes are likely to be affected more or less 
equally by deviations from Eq.~(\ref{fNLmodel}).

As we discuss below in greater detail, the parameters in our likelihood analysis include $\fNL$, the 
matter density $\Omega_m$ and the matter fluctuation amplitude $\sigma_8$, while we consider both a constant
 and time-varying dark energy equations of state described in terms of one ($w$) and two 
\citep[$w_0$ and $w_a$,][]{Chevallier:2000qy,Linder:2002et} parameters respectively.  For definiteness, we 
assume a fiducial ideal survey similar in size and depth to that of the upcoming South Pole Telescope/Dark 
Energy Survey \citep[SPT-DES,][]{Ruhl:2004kv,Abbott:2005bi}. We assume a $\Lambda$CDM fiducial cosmology, 
for two values of $\sigma_8$, since cluster number counts are extremely sensitive to this parameter. 

This paper is organized as follows. In section~\ref{secModel} we introduce our model for the non-Gaussian 
mass function and describe our analysis of the dependence of the expected errors on cosmological parameters 
on the non-Gaussian component. In section~\ref{secResults} we present our results and we conclude in 
section~\ref{secConclusions}.

\section{The model}
\label{secModel}

In this section we present the methods applied in the present work. We begin with a brief review of 
previous works dealing with non-Gaussian initial conditions in galaxy cluster observations, 
and then we describe in detail our treatment of the non-Gaussian mass function. We conclude this
section with a discussion of the likelihood analysis whose results will be given in 
section~\ref{secResults}.

\subsection{Historical overview}

Expressions for the cluster mass function in the presence of non-Gaussian initial conditions have been 
derived as extensions to the Press-Schechter ansantz \citep[PS,][]{Press:1973iz} first by 
\citet{Lucchin:1987yv} and \citet{Colafrancesco:1989px} while a simpler approach has been adopted 
later by \citet{Chiu:1997xb} and \citet{Robinson:1998dx}. 

The original PS formula describes the comoving number density $n(M)dM$ of clusters with mass in the 
interval $(M,M+dM)$ as 
\beq\label{nPS}
n_{PS}(M) dM = -\frac{2\bar{\rho}}{M}\frac{d}{dM}
\left[\int_{\delta_c / \sigma_M}^{\infty} P_G(y)dy\right] d M,
\eeq
where we suppress, for clarity, the redshift dependence, $\bar{\rho}$ is the comoving mass density, 
$\sigma_M$ is the r.m.s. of mass fluctuations in spheres of radius $R=(3M/4\pi\bar{\rho})^{1/3}$,  
$\delta_c=1.686$ is the critical linear overdensity in the spherical collapse model and $P_G$ is the 
Gaussian probability distribution function (PDF), $P_G(y)=e^{-y^2/2}/\sqrt{2\pi}$. Since the function 
$P_G(y)$ does not depend explicitly on the mass $M$, and therefore on the scale $R$, Eq.~(\ref{nPS}) 
reduces to
\beq\label{nPS2}
n_{PS}(M) dM = -\frac{2\bar{\rho}}{M^2}\frac{\delta_c}{\sigma_M}
\frac{d\ln\sigma_M}{d\ln M} P_G(\delta_c/\sigma_M) d M.
\eeq

The PS formalism assumes that the scale dependence of the PDF of the density field is completely described 
by the scale dependence of the variance $\sigma_M^2$. \citet{Lucchin:1987yv}, \citet{Colafrancesco:1989px} 
and, later \citet{Matarrese:2000iz} considered a derivation of the non-Gaussian mass function, based on 
Eq.~(\ref{nPS}), that takes into account the scale dependence of higher order cumulants, thereby allowing 
for a generic dependence of the PDF on the smoothing scale $R$. Specifically, \citet{Matarrese:2000iz} 
(hereafter MVJ) derived the mass function corresponding to the model described by Eq.~(\ref{fNLmodel}). 
The non-Gaussianity of the mass function is described, in first approximation, in terms of the 
skewness $S_{3,R}$ of the smoothed density field $\delta_R$,
\beq
S_{3,R}\equiv\frac{\lan \delta_R^3 \ran}{\lan \delta_R^2 \ran^2},
\eeq
and it is obtained from the cumulant generator of the distribution as 
\bea\label{nMVJ}
n_{MVJ}(M)dM\simeq -\frac{2\bar{\rho}}{M^2}\frac{1}{\sigma_M}\times\qquad \nonumber\\
\times\left[\frac{1}{2}\frac{\delta_c^3}{\delta_*}\frac{d S_{3,R}}{d\ln M}
+\delta_*\frac{d\ln\sigma_M}{d\ln M}\right]\frac{e^{-\delta_*^2/(2\sigma_M^2)}}{\sqrt{2\pi}}dM,
\eea
where $\delta_*=\delta_c\sqrt{1-S_{3,R}\delta_c/3}$.

It's worth noticing here that although Eq.~(\ref{fNLmodel}) should be seen as a truncated expansion in 
powers of $\phi$, the mass function provided by Eq.~(\ref{nMVJ}) is not linear in the non-Gaussian 
parameter $\fNL$ (since $S_{3,R}\sim \fNL$); rather it describes the non-Gaussian PDF by its proper 
dependence on the skewness while neglecting all higher order cumulants.

The simpler extension to non-Gaussian initial conditions introduced by \citet{Chiu:1997xb} consists 
instead of replacing the Gaussian function $P_G(y)$ in Eq.~(\ref{nPS2}) by the appropriate, 
non-Gaussian PDF $P_{NG}(y)$, assumed to be scale-independent. The resulting mass function, which 
we will denote here as ``extended-PS''  or EPS, therefore reads
\bea\label{nEPS}
n_{EPS}(M) dM = -\frac{2\bar{\rho}}{M^2}\frac{\delta_c}{\sigma_M}\times\nonumber\\
\times\frac{d\ln\sigma_M}{d\ln M} P_{NG}(\delta_c/\sigma_M) d M.
\eea

This approach, has been tested in N-body simulations by \citet{Robinson:1999se} for several non-Gaussian 
models; they  find that Eq.~(\ref{nEPS}) agrees with measurement of the cumulative mass function $n(>M)$
in the simulations to within $25\%$. While this error is slightly larger than the differences between 
the PS formula, Eq.~(\ref{nPS}), and simulation results for Gaussian initial conditions, it is much smaller
than the model-to-model differences between the cumulative mass functions. As a measure of the 
non-Gaussianity of the tail of the distribution function $P_{NG}(y)$, \citet{Robinson:1998dx} introduced the
parameter $G$ (there called $T$) defined as 
\beq\label{defG}
G=\frac{\int_3^\infty P_{NG}(y)dy}{\int_3^\infty P_G(y) dy}
\eeq
with $G=1$ corresponding to the Gaussian case. 

Following this approach \citet{Robinson:1998dx}, \citet{Koyama:1999fc} and \citet{Willick:1999ty} placed 
constraints on primordial non-Gaussianity from X-ray cluster survey observations \citep{Henry:1991ce,
Ebeling:1996zp,Henry:1997ic} and \citet{Amara:2003qt} relate primordial non-Gaussianity with the 
normalization of the dark matter power spectrum. In particular, assuming that the non-Gaussian primordial 
field can be generically described by a log-normal distribution, \citet{Robinson:1999wh} found, for a 
$\Lambda$CDM cosmology, the constraint $G<6$ at $2\sigma$ level. An analysis of the constraining power of 
future Sunyaev-Zel'dovich (SZ) cluster surveys on cosmological parameters which includes the possibility of 
primordial non-Gaussianity is provided by \citet{Benson:2001hc}. Specifically, this work assumes the 
log-normal PDF studied by \citet{Robinson:1999wh} and  performs a Fisher-matrix analysis that includes 
the matter and baryon density parameters $\Omega_m$ and $\Omega_b$, $\sigma_8$ and the non-Gaussian 
factor $G$. The results for the 1-$\sigma$ errors on $G$, assuming priors from CMB, Large-Scale 
Structure (LSS) and supernovae (SN) observations, are $\Delta G\simeq 2$ and $\Delta G\simeq 0.1$ for 
the Bolocam and Planck experiments respectively. 

Finally, \citet{Sadeh:2006ah} apply the same extended PS formalism to the $\chi_m^2$ non-Gaussian model 
\citep{White:1998da,Koyama:1999fc}. Here, however, much attention is devoted to highly non-Gaussian models, 
e.g., with $m=1$ and $2$, which are already excluded by measurements of the galaxy bispectrum in the PSCz 
survey \citep{Feldman:2000vk}.

\subsection{The non-Gaussian mass function}

In our analysis we will make use of the EPS approach, Eq.~(\ref{nEPS}), since it can be more easily 
implemented (once the probability function $P_{NG}(y)$ is known) and avoids problems with small regions of 
the parameter space where the MVJ expression for the mass function, Eq.~(\ref{nMVJ}), is beyond its limits 
of validity. For most of the cases considered in section~\ref{secResults}, however, we performed the 
analysis using both approaches, finding almost identical results. 

Since the PS and EPS expressions are known to differ by up to $25\%$ from N-body results, we use the EPS 
non-Gaussian mass function only to model {\it departures} from the Gaussian case; for the latter we use an
analytic mass function fit to the N-body results. Specifically, we consider the non-Gaussian mass function 
$n(z,M,\fNL)$ to be given by the product 
\beq\label{massfunction}
n(z,M,\fNL)=n_G(z,M)\;F_{NG}(z,M,\fNL),
\eeq
where $n_G(z,M)$, corresponding to the Gaussian case, is the fit to N-body simulations provided by 
\citet{Jenkins:2000bv},
\bea
n_G(z,M)dM=-0.301\frac{\rho_m}{M\sigma_M}\frac{d\sigma_M}{dM} \times\nonumber\\
\exp\left[-|0.64-\log[D(z)\sigma_M]|^{3.82}\right],
\eea
where $D(z)$ is the linear growth factor computed by solving the differential equation governing structure 
evolution. The non-Gaussian factor $F_{NG}(z,M,\fNL)$ is derived from the EPS mass function and simply given
by 
\beq
F_{NG}(z,M,\fNL)\equiv\frac{n_{EPS}(z,M,\fNL)}{n_{PS}(z,M)} 
\eeq
where $n_{PS}$ is the Gaussian PS mass function. Note that for $\fNL=0$ we have $n_{EPS}=n_{PS}$. 

\begin{figure}[t]
\begin{center}
\includegraphics[width=0.48\textwidth]{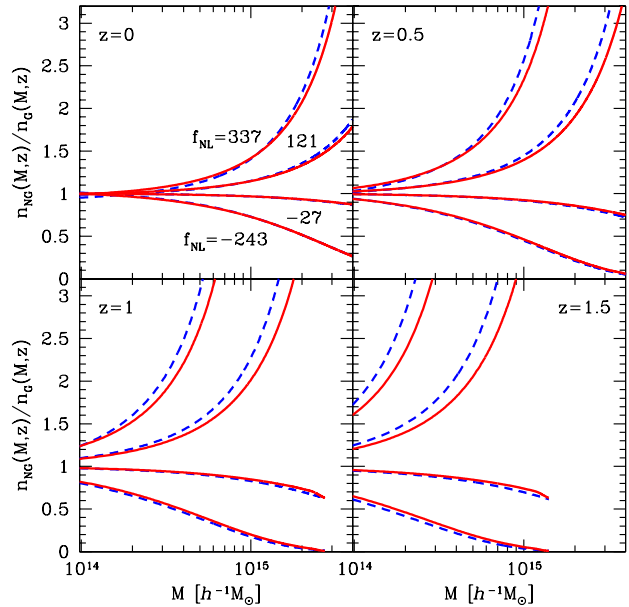} 
\caption{\label{Fig_ng_mass_func} Uncertainty on the mass function $n(M,z)$ due to non-Gaussianity expressed
as the ratio between the non-Gaussian to the Gaussian mass function at four different redshifts ($0$, $0.5$,
$1$ and $1.5$) for different values of $\fNL$. The inner continuous lines corresponds to the 
(2-$\sigma$) limits $-27<\fNL<121$ derived from the WMAP (first year) constraints on $\fNL$ by 
\citet{Creminelli:2005hu} while the outer continuous lines corresponds to the limits $-243<\fNL<337$ from the 
expected SDSS galaxy bispectrum constraints \citep{Scoccimarro:2003wn} computed with the EPS approach, 
Eq.~(\ref{nEPS}). The dashed lines, almost coincident with the continuous ones,
correspond to the same limits computed by means of the MVJ formula, Eq.~(\ref{nMVJ}).}
\end{center}
\end{figure}
It can be easily shown that the predictions of the EPS and MVJ methods are very close by comparing them for 
relevant values of the parameter $\fNL$. In Fig.~\ref{Fig_ng_mass_func} we plot the ratio of the 
non-Gaussian mass function to the Gaussian one at different redshifts and as a function of the mass $M$ for 
the 2-$\sigma$ limits 
\bdm
-27 < \fNL < 121,
\edm
obtained from the bispectrum analysis of the WMAP 1-year data by \citet{Creminelli:2005hu}, yielding 
constraints that are slightly tighther than but consistent with those obtained from the WMAP 3-year data 
by \citet{Spergel:2006hy}. We plot as well the limits
\bdm
-243 < \fNL < 337,
\edm
corresponding to the 1-$\sigma$ error $\dfNL=145$ expected from measurements of the galaxy bispectrum in the
Sloan Digital Sky Survey (SDSS) main sample \citep{Scoccimarro:2003wn}. Notice that in the latter case we 
are assuming here as fiducial value for the non-Gaussian parameter $\fNL=47$, i.e. the maximum 
likelihood value of the cited WMAP analysis. The continuous lines denote therefore the allowed region 
computed by means of the EPS formula, Eq.~(\ref{nEPS}), while the dashed lines corresponds to the same 
quantity determined assuming the EPS formalism of Eq.~(\ref{nEPS}). 

The non-Gaussian PDF $P_{NG}(y)$ that appears in the EPS formula, Eq.~(\ref{nEPS}), is measured from 
realizations on a $512^{3}$ grid in a box of $1\Gpc$ side of the curvature perturbation $\Phi$ described 
by Eq.~(\ref{fNLmodel}) in terms of the Gaussian field $\phi$, generated in Fourier space with a scale 
invariant power spectrum. The field $\Phi$ is then converted into the mass density field 
in Fourier space by means of Poisson equation and a transfer function computed by the CMBFAST code 
\citep{Seljak:1996is}, smoothed on a $R=4\Mpc$ scale and the probability distribution is finally measured 
in position space. We determine a probability function for a set of values of $\fNL$ from $-500$ 
to $500$ and then interpolated to the desired value. We assume, in all cases, the fiducial cosmology 
described below. In order to reliably estimate the tail of the distribution, several thousands realizations 
were needed for each value of $\fNL$. For high values of the mass $M$ and redshift, corresponding to 
extremely rare events (7-$\sigma$), the probability distributions could not be properly determined, as can 
be seen from the limited plots in the lower panels in Fig.~\ref{Fig_ng_mass_func}. None of the results in 
the paper, however, is sensitive to this cut-off.

It is evident from the figure that the difference between the two approaches, for this non-Gaussian model, 
is small, essentially noticeable just for large values of $\fNL$. This is due to the relatively mild 
dependence on the smoothing scale $R$ of the {\it reduced} skewness $s_{3,R}\equiv S_{3,R}\sigma_R$ for 
our non-Gaussian model, as we tested as well by choosing different smoothing lengths for the probability 
distributions measured from the realizations. On the other hand, the close results obtained by the different 
methods show how the degree of non-Gaussianity currently allowed can be described by the first moments of the 
primordial distribution, if not by the skewness alone. It is worth stressing, however, that both 
prescriptions for the non-Gaussian mass function, even when limited to modeling deviations from the 
Gaussian case, need to be properly tested against N-body simulations. New results in this direction will 
soon be available \citep{Matarrese:privcom}. 

By means of the mentioned measured probability functions, in the framework of the EPS approach, it is also 
possible to translate the constraints on the parameter $\fNL$ into constraints on the parameter $G$ 
defined in Eq.~(\ref{defG}). As an example, the WMAP 1-$\sigma$ error $\dfNL=37$ \citep{Creminelli:2005hu} 
corresponds to $\Delta G\simeq 0.06$ while the expected 1-$\sigma$ error $\dfNL=145$ from SDSS galaxy 
bispectrum measurement corresponds to $\Delta G\simeq 0.25$.

\subsection{Likelihood analysis}

In this section we describe the likelihood analysis we use to obtain our results. We will consider two 
simple models depending on four and five parameters. In addition to $\fNL$ we consider the matter density 
parameter $\Omega_m$, fluctuation amplitude parameter $\sigma_8$ and we will separately consider the cases
of dark energy with either a constant equation of state parameter ($w$) or a time-varying equation of state
described by two parameters ($w_0$ and $w_a$). In all cases we assume a spatially flat cosmological model 
for simplicity.

The fiducial values assumed for the likelihood analysis are given in Table~\ref{Tab_fid_val}. 
Since the expected number of observable clusters is highly dependent on the value of $\sigma_8$, for the 
four parameter model we perform the analysis assuming as well the lower value $\sigma_8=0.75$, while in 
every other case we assume $\sigma_8=0.9$. The choice of the fiducial value $\fNL=47$ for the non-Gaussian 
parameter does not substantially affect any of the results of the present work. 

Unless otherwise stated, we consider an ideal survey with limiting mass $\Mlim=1.75\times 10^{14}\Ms$ and 
with a sky coverage of $4000$ deg$^2$ ($f_{sky}\simeq 10\%$) out to a maximum cluster redshift of $1.5$, 
corresponding to the expectations for the SPT and DES projects. For our fiducial model with $\sigma_8=0.9$ 
and $\fNL=47$, this yields a total of $21,000$ clusters in $15$ redshift bins; if we had instead chosen 
$\fNL=0$ for the fiducial model, we would obtain $20,000$ clusters, consistent with earlier estimates 
\citep{Wang:2004pk}. 

We study the dependence on cosmology and on the constant $\fNL$ of the total number and mass distribution 
of clusters above a certain {\it fixed}, i.e. redshift independent, threshold mass $\Mlim$ and explore
the degeneracies introduced by varying non-Gaussian initial conditions. While the redshift dependence of 
the threshold mass should be included when making precise predictions for a given survey, this dependence 
is weak for SZ-selected cluster samples; as a result, our neglect of such dependence here will not 
significantly affect our conclusions. 

\begin{table}[t]
\caption{\label{Tab_fid_val} Fiducial values for the cosmological and non-Gaussian parameters.}
\begin{ruledtabular}
\begin{tabular}{llcc}
\multicolumn{2}{l}{Parameter}                         &  Fiducial value \\
\hline
$\Omega_m$       & matter density                     & $0.27$          \\
$\sigma_8$       & galaxy fluctuation amplitude       & $0.9$~($0.75$)  \\
$w$/$w_0$        & dark energy equation of state      & $-1$            \\
$w_a$            & dark energy equation of state      & $0$             \\
$\fNL$           & non-Gaussian parameter             & $47$            \\
\hline
$n_s$            & scalar spectral index              & $1$             \\
$h$              & Hubble parameter                   & $0.72$          \\
$\Omega_b h^2$   & physical baryon density            & $0.0232$        \\
$\Omega_\Lambda$ & dark energy density                & $1-\Omega_m$    \\
\end{tabular}
\end{ruledtabular}
\end{table}

The total number of clusters with mass $M$ above $\Mlim$, per unit redshift, is given by
\beq
\frac{dN}{dz}=\Delta\Omega\frac{dV}{dzd\Omega}(z)\int_{\Mlim}^\infty n(z,M,\fNL)dM
\eeq
where 
\beq
\frac{dV}{dzd\Omega}(z)=\frac{1}{H(z)}\left[\int_0^z\frac{dz'}{H(z')}\right]^2
\eeq
is the cosmology-dependent volume factor for flat models and $\Delta\Omega$ is the solid angle subtended by 
the survey area.

\begin{figure}[t]
\begin{center}
\includegraphics[width=0.48\textwidth]{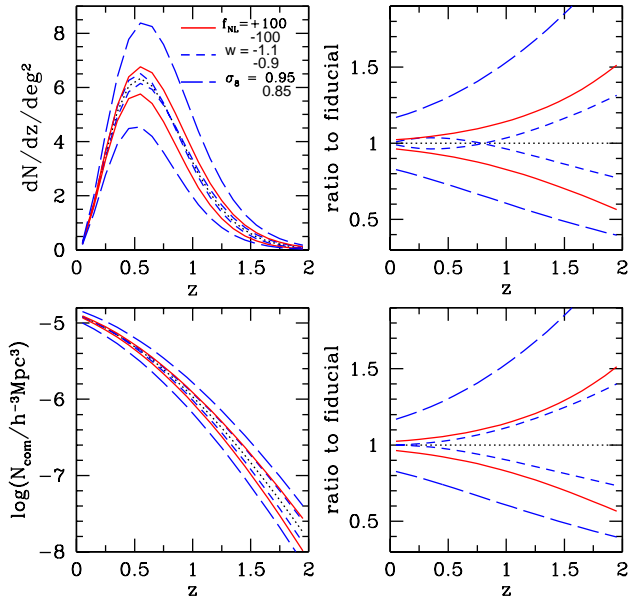}
\caption{\label{Fig_fNL_vs_w_s8} Cluster counts per unit redshift (upper panels) and comoving cluster 
density (lower panels) as a function of redshift for different values of the non-Gaussian parameter $\fNL$ 
($\fNL=\mp 100$, continuous lines), of the dark energy equation of state parameter $w$ ($w=-1.1$ and 
$-0.9$, short-dashed lines) and of $\sigma_8$ ($\sigma_8=0.85$ and $0.95$, long-dashed lines) compared to 
the fiducial case (dotted line) with $\sigma_8=0.9$, $w=-1$ and $\fNL=0$. Assumes the mass limit 
$\Mlim=1.75\times 10^{14}\Ms$.}
\end{center}
\end{figure}

We show in  Fig.~\ref{Fig_fNL_vs_w_s8} the sensitivity of the {\it total} number of clusters above 
$\Mlim$ per unit redshift and unit area (upper panels) and of the comoving number density (lower 
panels) to different values of the non-Gaussian parameter $\fNL$ compared with the sensitivity to different 
values of the dark energy equation of state parameter $w_0$ and of the fluctuation amplitude parameter 
$\sigma_8$. 

Fig.~\ref{Fig_fNL_vs_w_s8} (upper left panel) shows that varying $\fNL$ over the range allowed by current
CMB observations yields changes in the cluster counts comparable to a $10\%$ variation in the dark energy 
equation of state parameter $w$. However, the upper right panel of Fig.~\ref{Fig_fNL_vs_w_s8} shows that 
the redshift dependence of the mass function variations due to non-Gaussianity are different from the 
variations due to changes in $w$. This is essentially due to the fact that $w$ affects both the mass 
function and the volume factor. On the other hand, the redshift dependence of variations due to changes in 
$\fNL$ appears more similar to variations induced by changes in $\sigma_8$, so we expect a stronger 
degeneracy between these two parameters.

\begin{figure}[t]
\begin{center}
\includegraphics[width=0.48\textwidth]{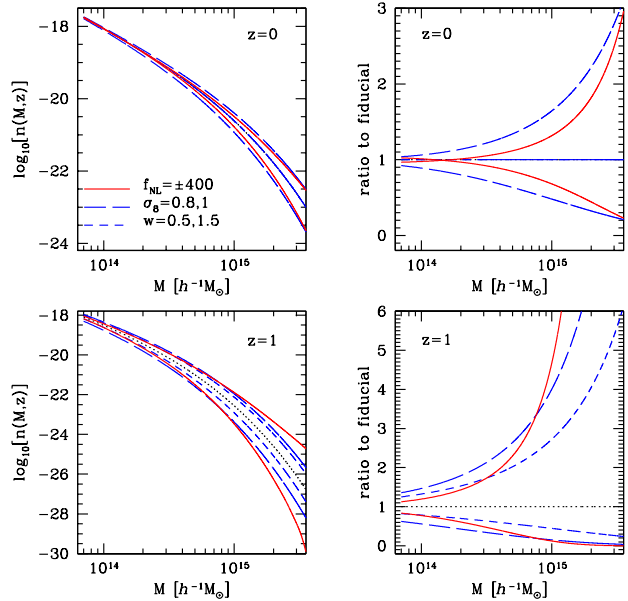}
\caption{\label{Fig_n_vs_M} The mass function $n(M,z)$ as a function of the mass $M$ at $z=0$ 
(upper panels) and $z=1$ (lower panels) for different values of the non-Gaussian parameter $\fNL$ 
(continuous lines), of the dark energy equation of state parameter $w_0$ (short-dashed lines) and of 
$\sigma_8$ (long-dashed lines) compared to the fiducial case (dotted line) with $\sigma_8=0.9$, $w=-1$ 
and $\fNL=0$.}
\end{center}
\end{figure}

In Fig.~\ref{Fig_n_vs_M} we show the sensitivity of the mass function $n(M,z)$ on the same parameters, this 
time as a function of the mass $M$ for $z=0$ (upper panels) and $z=1$ (lower panels). In this case the 
behavior of the cluster density as we vary $\fNL$ and $\sigma_8$ is quite different. One can clearly see 
how non-Gaussianity is particularly significant for the high-mass tail of the distribution. This fact 
suggests that it might be relevant to consider a likelihood analysis that takes into account the {\it full} 
functional shape of the mass function by dividing the observable clusters in mass bins \citep[see, for 
instance,][]{Hu:2003sh,Lima:2005tt,Marian:2006zp}. In this way one might expect to mitigate the 
$\fNL$-$\sigma_8$ degeneracy evident from Fig.~\ref{Fig_fNL_vs_w_s8} and better study the possibility of 
constraining non-Gaussianity with cluster surveys.

In the next section we will consider the two cases of an analysis involving a single mass bin defined by 
$M>\Mlim$ and of an analysis with several mass bins. The likelihood function is based on the 
assumption of Poisson statistics for the cluster number measurements in each redshift bin, so that, for the 
single mass bin case we have
\beq
\ln {\mathcal L}=\sum_{i=1}^{N_{tot}^z}\left[N_i\ln N_i^*-N_i^*-N_i\ln N_i+N_i\right]
\eeq
where $N_{tot}^z$ is the total number of redshift bins, $N_i^*$ is the fiducial number count in the 
$i$-th redshift bin, $N_i$ is the number count of clusters in the $i$-th redshift bin for the specific 
model. Throughout the paper, we consider $N_{tot}^z=15$ redshift bins with width $\Delta z=0.1$ out to 
$z=1.5$.

In the case of multiple mass bins, the likelihood function is of the form
\bea\label{likelihoodmass}
\ln {\mathcal L}=\sum_{i=1}^{N_{tot}^z}\sum_{j=1}^{N_{tot}^M}
\left[N_{ij}\ln N_{ij}^*-N_{ij}^*\right.\nonumber \\
\left.-N_{ij}\ln N_{ij}+N_{ij}\right]
\eea
with $N_{tot}^M=10$ being the total number of mass bins logarithmically spaced from $\Mlim$ to 
$M_{\rm max}=5\times 10^{15}\Ms$; here $N_{ij}$ is the number of model clusters in the $i$-th redshift bin 
and $j$-th mass bin.

The marginalization of the likelihood functions is performed on a regular grid with a varying number of 
points chosen to optimize the sampling of the parameter space.

We do not include systematic errors which a real survey will encounter, including uncertainties in the 
cluster mass-observable relation \citep[e.g.][]{Seljak:2001cj,Levine:2002uq,Pierpaoli:2002rh,Majumdar:2002hd,
Majumdar:2003mw,Hu:2003sh,Lima:2004wn,Francis:2005nu,Lima:2005tt,Kravtsov:2006db} and in cluster redshift 
determination \citep[e.g.][]{Huterer:2004rf}. We excluded as well statistical uncertainties related to 
sample variance \citep[e.g.][]{Hu:2002we} or theoretical uncertainties in the cluster mass function and 
its cosmological dependence \citep[e.g.][]{Heitmann:2004gz,Warren:2005ey,Reed:2006rw,Crocce:2006ve}. Finally,
we note that other degeneracies with parameters such as those describing spatial curvature \citep{Abbott:2005bi} 
and the effect of massive neutrinos on the dark matter power spectrum \citep{Huterer:2006mv} might be relevant 
for future high precision analyses.

\section{Results}
\label{secResults}

In this section we estimate the impact of marginalizing over the non-Gaussian parameter $\fNL$ on the 
determination of the dark energy equation of state as well as on two other relevant cosmological parameters 
such as the matter density $\Omega_m$ and fluctuation amplitude $\sigma_8$. We will separately consider the 
case of a dark energy equation of state determined by a single parameter ($w$) and the case of a 
two-parameter description of a time-varying equation of state ($w_0$ and $w_a$).

We derive the marginalized errors on the parameters with fixed $\fNL=47$ (no marginalization) and with 
three different Gaussian priors on $\fNL$, two corresponding to the constraints from CMB bispectrum 
measurements expected from the Planck experiment \citep{Komatsu:2001rj,Liguori:2005rj} and measured in the 
WMAP experiment \citep{Creminelli:2005hu} with
\beq
\fNL = 47\pm 5 \qquad ({\rm 1-}\sigma,~{\rm Planck})
\eeq
and
\beq
\fNL = 47\pm 37 \qquad ({\rm 1-}\sigma,~{\rm WMAP}),
\eeq
and a third corresponding to the expected constraints from the analysis of the SDSS main sample 
galaxy bispectrum \citep{Scoccimarro:2003wn},
\beq
\fNL=47\pm 145 \qquad ({\rm 1-}\sigma,~{\rm SDSS~ forecast}).
\eeq
This last case is motivated by a possible strong scale-dependence of primordial non-Gaussianity, not 
captured by the model defined by Eq.~(\ref{fNLmodel}), that could result in a stronger non-Gaussian effect
at smaller scales, thereby escaping the CMB constraints. As a rough estimate of the smallest scale probed by
the mentioned experiments, we notice that for WMAP, the maximum multipole $l_{max}\simeq 1000$ corresponds
to $\sim 50\Mpc$ while Planck is expected to probe a scale three times smaller; in the SDSS case a maximum 
comoving wavenumber $k_{max}\simeq 0.3 \kMpc$ corresponds to $20\Mpc$. The typical scale probed by clusters is about $5$ to $10\Mpc$, with the most massive clusters approaching the lowest scale probed by Planck. 

In all the different cases considered we include as well the results obtained with two, independent, 
Gaussian priors on $\Omega_m$ and $\sigma_8$ with errors roughly corresponding to the knowledge provided by 
WMAP observations for a $\Lambda$CDM model \citep{Spergel:2006hy} in combination with other probes, such as, 
for example, the LSS power spectrum,
\beq
\sigma_8 = 0.9\pm 0.05, \qquad{\rm and}\qquad \Omega_m=0.27\pm 0.035,
\eeq
and by future constraints from Planck in combination with other probes \citep{unknown:2006uk},
\beq
\sigma_8 = 0.9\pm 0.01, \qquad{\rm and}\qquad \Omega_m=0.27\pm 0.0035.
\eeq
As an extreme example, in the last two lines of tables, we give results corresponding to fixing $\Omega_m$ 
and $\sigma_8$, studying therefore a likelihood function for the dark energy parameters and $\fNL$ alone. 

We caution that these priors have a purely illustrative significance and are chosen here for the sake of 
simplicity. A proper treatment of external data sets, which is beyond the scope of this paper, would 
naturally involve the parameters covariance and it would affect directly the dark energy parameters as 
well. On the other hand, even rigorous analyses of CMB or LSS galaxy power spectra would probably be 
insensitive to the non-Gaussian parameter $\fNL$.

\subsection{1-parameter Dark Energy equation of state}

\squeezetable
\begin{table}[t]
\caption{\label{Table_1DE} Expected cosmological errors ($1-\sigma$) from the cluster survey for the 
4-parameter ($\Omega_m$, $\sigma_8$, $w$, $\fNL$) analysis. The percentages in parentheses express the 
increase in the error with respect to the case without marginalization on $\fNL$ ($\dfNL=0$). We assume 
a fiducial $\sigma_8=0.9$ and one mass bin defined by $M>\Mlim=1.75\times 10^{14}\Ms$.}
\begin{ruledtabular}
\begin{tabular}{l|cccc}
prior:            & $\dfNL=0$  & $\dfNL=5$       & $\dfNL=37$         & $\dfNL=145$        \\
\hline\hline
\multicolumn{5}{l}{No priors on $\Omega_m$ and $\sigma_8$}                                 \\
\hline
$\Delta w$        & $0.045$  & $0.045$~($0\%$)   & $0.049$~($9\%$)    & $0.079$~($76\%$)   \\
$\Delta \Omega_m$ & $0.0085$ & $0.0085$~($0\%$)  & $0.0085$~($0\%$)   & $0.0087$~($2\%$)   \\
$\Delta \sigma_8$ & $0.0051$ & $0.0052$~($2\%$)  & $0.0083$~($63\%$)  & $0.0226$~($340\%$) \\
$\dfNL$           & -        & $5.0$             & $37$               & $123$              \\
\hline\hline
\multicolumn{5}{l}{Gaussian priors: $\Omega_m=0.27\pm 0.035$, $\sigma_8= 0.9\pm 0.05$}     \\ 
\hline
$\Delta w$        & $0.044$  & $0.044$~($0\%$)   & $0.048$~($9\%$)    & $0.076$~($73\%$)   \\
$\Delta \Omega_m$ & $0.0082$ & $0.0082$~($0\%$)  & $0.0082$~($0\%$)   & $0.0083$~($1\%$)   \\
$\Delta \sigma_8$ & $0.0050$ & $0.0050$~($0\%$)  & $0.0081$~($62\%$)  & $0.0205$~($310\%$) \\
$\dfNL$           & -        & $5.0$             & $36$               & $113$              \\
\hline\hline
\multicolumn{5}{l}{Gaussian priors: $\Omega_m=0.27\pm 0.0035$, $\sigma_8= 0.9\pm 0.01$}    \\ 
\hline
$\Delta w$        & $0.023$  & $0.024$~($4\%$)   & $0.031$~($35\%$)   & $0.042$~($83\%$)   \\
$\Delta \Omega_m$ & $0.0032$ & $0.0032$~($0\%$)  & $0.0032$~($0\%$)   & $0.0032$~($0\%$)   \\
$\Delta \sigma_8$ & $0.0021$ & $0.0023$~($10\%$) & $0.0055$~($160\%$) & $0.0091$~($330\%$) \\
$\dfNL$           & -        & $5.0$             & $31$               & $54$               \\
\hline\hline
\multicolumn{5}{l}{Fixed $\Omega_m=0.27$ and $\sigma_8=0.9$}                               \\
\hline
$\Delta w$        & $0.0172$ & $0.0177$~($3\%$)  & $0.0184$~($9\%$)   & $0.0184$~($9\%$)   \\
$\dfNL$           & -        & $3.8$             & $5.6$              & $5.7$              \\
\end{tabular}
\end{ruledtabular}
\end{table}
The main results in this paper are shown in Table~\ref{Table_1DE}, where we present the expected 1-$\sigma$ 
errors from the cluster survey for the three parameters $\Omega_m$, $\sigma_8$ and $w$ with no 
marginalization on $\fNL$ ($\dfNL=0$) and with a marginalization that includes the three Gaussian priors 
discussed above ($\dfNL=5$, $37$ and $145$), assuming in this case a fiducial $\sigma_8=0.9$. The 
percentages in parentheses express the increase in the error with respect to the case without 
marginalization on $\fNL$. Although the derived marginalized likelihood for single parameters are quite
close to Gaussian functions, the estimated errors reported in Table~\ref{Table_1DE} as in the following ones, 
are given, for clarity, by the mean between upper and lower errors.

\begin{figure*}[t]
\begin{center}
\includegraphics[width=0.98\textwidth]{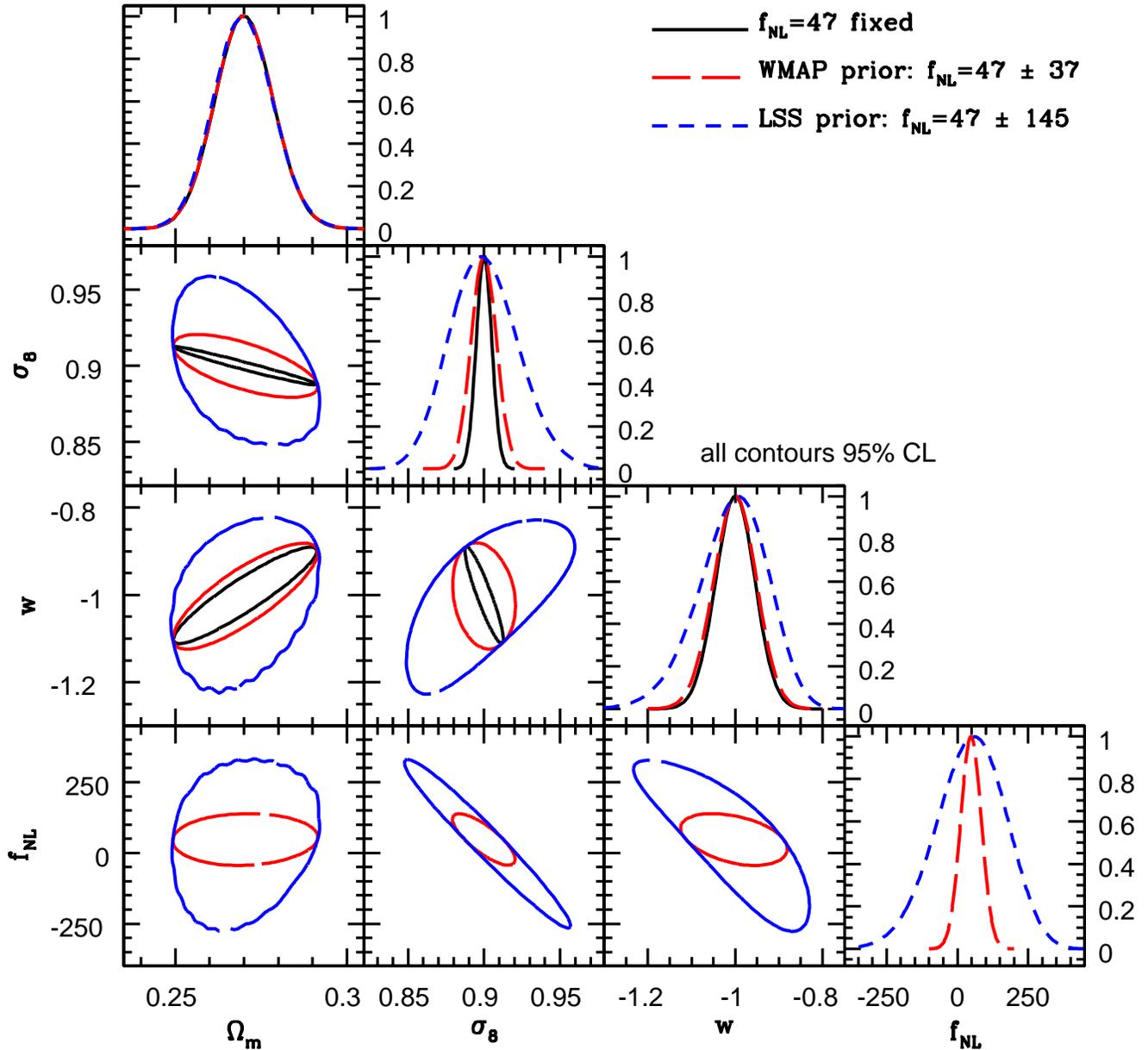}
\caption{\label{Fig_contours_3P} Forecast marginalized likelihoods and $95\%$ C.L. contour plots from the 
cluster survey for the 4-parameter ($\Omega_m$, $\sigma_8$, $w$ and $\fNL$) model. In each contour plot, 
the other two parameters are marginalized over; in each likelihood plot, the other three parameters are 
marginalized over. We assume a fiducial $\Lambda$CDM model with $\sigma_8=0.9$ and use one mass bin defined 
by $M>\Mlim=1.75\times 10^{14}\Ms$.} 
\end{center}
\end{figure*}
The major conclusion from Table~\ref{Table_1DE} is that inclusion of a possible non-Gaussian component at
the level allowed by present and future CMB constraints will not have an appreciable impact on the 
determination of dark energy parameters from cluster surveys. On the other hand, using only a single 
cluster mass bin (i.e., no information about the shape of the cluster mass function) and the WMAP prior on 
$\fNL$, the inclusion of non-Gaussianity degrades the cluster constraint on $\sigma_8$ by $50\%$. 
Moreover, using only the projected SDSS bispectrum constraint on $\fNL$, we do see degeneracies between 
$\fNL$ and dark energy: the error on $w$ from clusters increases by $\sim 70\%$, and the error on $\sigma_8$ 
grows by a factor of more than three compared to the purely Gaussian case. In all cases, the determination of 
$\Omega_m$ stays largely unaffected. 

The expected degeneracy between $\fNL$ and $\sigma_8$ is evident from the two-parameter $95\%$ C.L. contour 
plots shown in Fig.~\ref{Fig_contours_3P}, where we marginalized over the remaining parameters. The same 
overall behavior is observed when we impose priors on $\Omega_m$ and $\sigma_8$. In 
Fig.~\ref{Fig_contours_3P}, the marginalized likelihood plot for $w$ shows that inclusion of the WMAP prior 
on $\fNL$ leads to essentially the same dark energy sensitivity for the cluster survey as one would have by 
fixing $\fNL$ (i.e., by not including non-Gaussianity).

\squeezetable
\begin{table}[t]
\caption{\label{Table_1DE_s80p75} Same as Table~\ref{Table_1DE} but with fiducial $\sigma_8=0.75$.}
\begin{ruledtabular}
\begin{tabular}{l|cccc}
prior:            & $\dfNL=0$ & $\dfNL=5$        & $\dfNL=37$         & $\dfNL=145$        \\
\hline\hline
\multicolumn{5}{l}{No priors}                                                              \\
\hline
$\Delta w$        & $0.079$   & $0.079$~($0\%$)  & $0.083$~($5\%$)    & $0.124$~($57\%$)   \\
$\Delta \Omega_m$ & $0.0140$  & $0.0140$~($0\%$) & $0.0140$~($0\%$)   & $0.0144$~($3\%$)   \\
$\Delta \sigma_8$ & $0.0076$  & $0.0076$~($0\%$) & $0.0100$~($32\%$)  & $0.0238$~($210\%$) \\
$\dfNL$           & -         & $5.0$            & $37$               & $128$              \\
\hline\hline
\multicolumn{5}{l}{Gaussian priors: $\Omega_m=0.27\pm 0.035$, $\sigma_8= 0.75\pm 0.05$}    \\ 
\hline
$\Delta w$        & $0.073$   & $0.073$~($0\%$)  & $0.079$~($8\%$)    & $0.119$~($63\%$)   \\
$\Delta \Omega_m$ & $0.0129$  & $0.0129$~($0\%$) & $0.0129$~($0\%$)   & $0.0129$~($0\%$)   \\
$\Delta \sigma_8$ & $0.0070$  & $0.0070$~($0\%$) & $0.0094$~($34\%$)  & $0.0212$~($200\%$) \\
$\dfNL$           & -         & $5.0$            & $37$               & $118$              \\
\hline\hline
\multicolumn{5}{l}{Gaussian priors: $\Omega_m=0.27\pm 0.0035$, $\sigma_8= 0.75\pm 0.01$}   \\ 
\hline
$\Delta w$        & $0.036$   & $0.036$~($0\%$)  & $0.046$~($28\%$)   & $0.063$~($75\%$)   \\
$\Delta \Omega_m$ & $0.0033$  & $0.0033$~($0\%$) & $0.0034$~($3\%$)   & $0.0034$~($3\%$)   \\
$\Delta \sigma_8$ & $0.0021$  & $0.0023$~($10\%$)& $0.0054$~($160\%$) & $0.0091$~($330\%$) \\
$\dfNL$           & -         & $5.0$            & $32$               & $56$               \\
\hline\hline
\multicolumn{5}{l}{Fixed $\Omega_m=0.27$ and $\sigma_8=0.75$}                              \\
\hline
$\Delta w$        & $0.030$   & $0.031$~($3\%$)  & $0.034$~($10\%$)   & $0.034$~($10\%$)   \\
$\dfNL$           & -         & $4.2$            & $7.5$              & $7.6$              \\
\end{tabular}
\end{ruledtabular}
\end{table}
As far as the constraints on non-Gaussianity are concerned, the information provided by the cluster 
likelihood on $\fNL$ adds little to that from the CMB, while it mildly improves upon the non-Gaussianity
constraints expected from the galaxy bispectrum. Even in the ideal case of perfect knowledge of $\Omega_m$ 
and $\sigma_8$, the expected error on $\fNL$ is $\dfNL\simeq 6$, of the same order of the expected error 
from Planck. 

We performed the same analysis for a fiducial $\sigma_8=0.75$, since this lower value has been recently 
suggested by CMB \citep{Spergel:2006hy} and cluster observations \citep{Gladders:2006uh,Dahle:2006fa}; 
as is well known, a lower clustering amplitude reduces the number of expected clusters and thereby reduces
the constraining power of cluster surveys. The results are given in Table~\ref{Table_1DE_s80p75}. In this 
case the total number of clusters for the fiducial model is about $6,000$; as a consequence, the 
cosmological constraints from the cluster survey are weaker than for the high-$\sigma_8$ model. However, the
relative impact of the marginalization over primordial non-Gaussianity is reduced. This result can be 
expected since the effect of imposing the {\it same priors} on $\fNL$ is relatively smaller when the 
cosmological errors on the other parameters for the fixed $\fNL$ case increase. 

To further illustrate the dependence of the results on survey parameters, in Table~\ref{Table_1DE_Mmin1p00} 
we show the constraints obtained when the threshold cluster mass is reduced to $\Mlim=1\times 10^{14}\Ms$. 
This lower threshold may be achieved, e.g., by supplementing SZ cluster detection with optical cluster 
selection using the red galaxy sequence \citep[e.g.][]{Gladders:2006uh,Koester:2006}. In this case, the 
$4,000$ deg$^2$ survey to $z=1.5$ includes about $75,000$ clusters, and the forecast cosmological parameter 
errors (without non-Gaussianity) are smaller by almost a factor of two than for the case with larger $\Mlim$
considered above. For this more sensitive cluster survey, the impact on cosmological parameters of 
marginalizing over $\fNL$ is correspondingly larger: while the impact on dark energy remains small, 
including non-Gaussianity with the WMAP prior expands the error on $\sigma_8$ by more than $100\%$.

As already discussed in the previous section, the degeneracy between $\sigma_8$ and $\fNL$ could be 
partially reduced by introducing a number of cluster mass bins and using the information contained in the 
shape of the mass function. In Table~\ref{Table_1DE_nb10}, we present the results for an analysis with a 
fiducial $\sigma_8=0.9$ and $\Mlim=1.75\times 10^{14}\Ms$ as in Table~\ref{Table_1DE} but subdividing the 
clusters into ten mass bins and using the likelihood function defined in Eq.~(\ref{likelihoodmass}). As the 
last column in Table~\ref{Table_1DE_nb10} indicates the main effect of including mass bins is that the 
cluster constraint on the non-Gaussian parameter $\fNL$ becomes stronger than that from the SDSS galaxy 
bispectrum. Even without combining with external data sets one can reach a $1-\sigma$ error of 
$\dfNL\simeq 50$, not too far from current limits from CMB observations. Further study would be needed to 
determine if this conclusion remains when realistic uncertainties in the cluster mass-observable relation 
are included in the analysis.

\subsection{2-parameter Dark Energy equation of state}

Finally we consider the case of a time-varying dark energy equation of state \citep{Chevallier:2000qy,
Linder:2002et}, 
\beq
w(a)=w_0+(1-a)w_a,
\eeq
adding the parameter $w_a$ to the likelihood analysis studied so far. In this case the strong degeneracy 
between $w_0$ and $w_a$ enlarges considerably the region of parameter space that has to be covered for the 
likelihood function evaluation, including unphysical regions where the combination $w_0+w_a$, representing 
the equation of state at large redshift, takes large positive values. To avoid such cases we impose, by 
hand, a Gaussian prior on the value of $\Omega_m(z)$, requiring in particular $1-\Omega_m(z)<0.01$  at 
$1$-$\sigma$ at $z=30$, the initial redshift considered for the numerical solution to the differential 
equation governing the growth factor $D(z)$. This ensures that the Universe is matter-dominated at early 
times as required by structure growth.

\squeezetable
\begin{table}[t]
\caption{\label{Table_1DE_Mmin1p00} Same as Table~\ref{Table_1DE} but 
with $\Mlim=1\times 10^{14}\Ms$.}
\begin{ruledtabular}
\begin{tabular}{l|cccc}
prior:            & $\dfNL=0$ & $\dfNL=5$        & $\dfNL=37$         & $\dfNL=145$        \\
\hline\hline
\multicolumn{5}{l}{No priors on $\Omega_m$ and $\sigma_8$}                                 \\
\hline
$\Delta w$        & $0.026$   & $0.026$~($0\%$)  & $0.030$~($15\%$)   & $0.052$~($100\%$)  \\
$\Delta \Omega_m$ & $0.0050$  & $0.0050$~($0\%$) & $0.0050$~($0\%$)   & $0.0052$~($4\%$)   \\
$\Delta \sigma_8$ & $0.0031$  & $0.0032$~($3\%$) & $0.0066$~($110\%$) & $0.0186$~($500\%$) \\
$\dfNL$           & -         & $5.0$            & $36$               & $113$              \\
\hline\hline
\multicolumn{5}{l}{Gaussian priors: $\Omega_m=0.27\pm 0.035$, $\sigma_8= 0.9\pm 0.05$}     \\ 
\hline
$\Delta w$        & $0.026$   & $0.026$~($0\%$)  & $0.030$~($15\%$)   & $0.050$~($92\%$)   \\
$\Delta \Omega_m$ & $0.0050$  & $0.0050$~($0\%$) & $0.0050$~($0\%$)   & $0.0051$~($2\%$)   \\
$\Delta \sigma_8$ & $0.0030$  & $0.0031$~($3\%$) & $0.0066$~($120\%$) & $0.0174$~($480\%$) \\
$\dfNL$           & -         & $5.0$            & $36$               & $106$              \\
\hline\hline
\multicolumn{5}{l}{Gaussian priors: $\Omega_m=0.27\pm 0.0035$, $\sigma_8= 0.9\pm 0.01$}    \\ 
\hline
$\Delta w$        & $0.017$   & $0.017$~($0\%$)   & $0.023$~($35\%$    & $0.032$~($88\%$) \\
$\Delta \Omega_m$ & $0.0028$  & $0.0028$~($0\%$)  & $0.0028$~($0\%$)   & $0.0028$~($0\%$)  \\
$\Delta \sigma_8$ & $0.0018$  & $0.0020$~($11\%$) & $0.0051$~($180\%$) & $0.0087$~($380\%$)\\
$\dfNL$           & -         & $5.0$             & $32$               & $56$              \\
\hline\hline
\multicolumn{5}{l}{Fixed $\Omega_m=0.27$ and $\sigma_8=0.9$}                               \\
\hline
$\Delta w$        & $0.0100$ & $0.0102$~($2\%$)  & $0.0102$~($2\%$)   & $0.0102$~($2\%$)   \\
$\dfNL$           & -        & $3.1$             & $4.0$              & $4.0$              \\
\end{tabular}
\end{ruledtabular}
\end{table}

\squeezetable
\begin{table}[b]
\caption{\label{Table_1DE_nb10} Same as Table~\ref{Table_1DE} but using 10 cluster mass bins.}
\begin{ruledtabular}
\begin{tabular}{l|cccc}
prior:            & $\dfNL=0$  & $\dfNL=5$       & $\dfNL=37$         & $\dfNL=145$        \\
\hline\hline
\multicolumn{5}{l}{No priors on $\Omega_m$ and $\sigma_8$}                                 \\
\hline
$\Delta w$        & $0.044$  & $0.044$~($0\%$)   & $0.046$~($5\%$)    & $0.048$~($9\%$)    \\
$\Delta \Omega_m$ & $0.0082$ & $0.0082$~($0\%$)  & $0.0083$~($1\%$)   & $0.0085$~($4\%$)   \\
$\Delta \sigma_8$ & $0.0049$ & $0.0050$~($2\%$)  & $0.0077$~($57\%$)  & $0.0103$~($110\%$) \\
$\dfNL$           & -        & $5.0$             & $29$               & $45$               \\
\hline\hline
\multicolumn{5}{l}{Gaussian priors: $\Omega_m=0.27\pm 0.035$, $\sigma_8= 0.9\pm 0.05$}     \\ 
\hline
$\Delta w$        & $0.043$  & $0.043$~($0\%$)   & $0.045$~($5\%$)    & $0.048$~($12\%$)   \\
$\Delta \Omega_m$ & $0.0079$ & $0.0079$~($0\%$)  & $0.0080$~($1\%$)   & $0.0082$~($4\%$)   \\
$\Delta \sigma_8$ & $0.0048$ & $0.0049$~($2\%$)  & $0.0075$~($56\%$)  & $0.0100$~($110\%$) \\
$\dfNL$           & -        & $5.0$             & $29$               & $45$               \\
\hline\hline
\multicolumn{5}{l}{Gaussian priors: $\Omega_m=0.27\pm 0.0035$, $\sigma_8= 0.9\pm 0.01$}    \\ 
\hline
$\Delta w$        & $0.023$  & $0.023$~($0\%$)   & $0.029$~($26\%$)   & $0.034$~($48\%$)   \\
$\Delta \Omega_m$ & $0.0032$ & $0.0032$~($0\%$)  & $0.0032$~($0\%$)   & $0.0032$~($0\%$)   \\
$\Delta \sigma_8$ & $0.0021$ & $0.0022$~($5\%$)  & $0.0048$~($130\%$) & $0.0063$~($200\%$) \\
$\dfNL$           & -        & $5.0$             & $26$               & $35$               \\
\hline\hline
\multicolumn{5}{l}{Fixed $\Omega_m=0.27$ and $\sigma_8=0.9$}                               \\
\hline
$\Delta w$        & $0.0166$ & $0.0173$~($4\%$)  & $0.0181$~($9\%$)   & $0.0181$~($9\%$)   \\
$\dfNL$           & -        & $3.7$             & $5.5$              & $5.5$              \\
\end{tabular}
\end{ruledtabular}
\end{table}

\squeezetable
\begin{table}[t]
\caption{\label{Table_2DE} Expected cosmological errors ($1\sigma$) from the cluster survey for the 
5-parameter ($\Omega_m$, $\sigma_8$, $w_0$, $w_a$, $\fNL$) analysis. We assume a fiducial $\sigma_8=0.9$ 
and one mass bin defined by $\Mlim=1.75\times 10^{14}\Ms$.}
\begin{ruledtabular}
\begin{tabular}{l|cccc}
prior:            & $\dfNL=0$ & $\dfNL=5$ & $\dfNL=37$ & $\dfNL=145$                        \\
\hline\hline
\multicolumn{5}{l}{No priors on $\Omega_m$ and $\sigma_8$.}                                 \\
\hline
$\Delta w_0$      & $0.195$   & $0.195$~($0\%$)  & $0.196$~($1\%$)    & $0.208$~($7\%$)     \\
$\Delta w_a$      & $0.73$    & $0.73$~($0\%$)   & $0.74$~($1\%$)     & $0.87$~($19\%$)     \\
$\Delta \Omega_m$ & $0.0156$  & $0.0156$~($0\%$) & $0.0158$~($1\%$)   & $0.0185$~($19\%$)   \\
$\Delta \sigma_8$ & $0.0086$  & $0.0087$~($1\%$) & $0.0114$~($32\%$)  & $0.0296$~($240\%$)  \\
$\dfNL$           & -         & $5.0$            & $37$               & $137$               \\
\hline\hline
\multicolumn{5}{l}{Gaussian priors: $\Omega_m = 0.27 \pm 0.035$, $\sigma_8 = 0.9 \pm 0.05$} \\ 
\hline
$\Delta w_0$      & $0.178$   & $0.178$~($0\%$)  & $0.178$~($0\%$)   & $0.183$~($3\%$)      \\
$\Delta w_a$      & $0.68$    & $0.68$~($0\%$)   & $0.68$~($0\%$)    & $0.73$~($7\%$)       \\   
$\Delta \Omega_m$ & $0.0140$  & $0.0140$~($0\%$) & $0.0141$~($1\%$)  & $0.0151$~($8\%$)     \\
$\Delta \sigma_8$ & $0.0078$  & $0.0078$~($0\%$) & $0.0105$~($35\%$) & $0.0240$~($210\%$)   \\
$\dfNL$           & -         & $5.0$            & $37$              & $118$                \\
\hline\hline
\multicolumn{5}{l}{Gaussian priors: $\Omega_m = 0.27 \pm 0.0035$, $\sigma_8 = 0.9 \pm 0.01$}\\ 
\hline
$\Delta w_0$      & $0.082$   & $0.082$~($0\%$)  & $0.85$~($4\%$)     & $0.89$~($9\%$)      \\
$\Delta w_a$      & $0.43$    & $0.43$~($0\%$)   & $0.43$~($0\%$)     & $0.43$~($0\%$)      \\   
$\Delta \Omega_m$ & $0.0034$  & $0.0034$~($0\%$) & $0.0034$~($0\%$)   & $0.034$~($0\%$)     \\
$\Delta \sigma_8$ & $0.0021$  & $0.0023$~($10\%$)& $0.0055$~($160\%$) & $0.0091$~($330\%$)  \\
$\dfNL$           & -         & $5.0$            & $31$               & $54$                \\
\hline\hline
\multicolumn{5}{l}{Fixed $\Omega_m=0.27$ and $\sigma_8=0.9$}                                \\
\hline
$\Delta w_0$      & $0.068$    & $0.069$~($1\%$) & $0.071$~($4\%$)    & $0.071$~($4\%$)     \\
$\Delta w_a$      & $0.37$     & $0.039$~($5\%$) & $0.40$~($8\%$)     & $0.40$~($8\%$)      \\
$\dfNL$           & -          & $3.9$           & $6.1$              & $6.2$               \\
\end{tabular}
\end{ruledtabular}
\end{table}
In Table~\ref{Table_2DE} we present the derived 1-$\sigma$ errors on the five parameters $\Omega_m$, 
$\sigma_8$, $w_0$ and $w_a$ with and without marginalization on $\fNL$ and assuming a single mass bin 
defined by $M>\Mlim=1.75\times 10^{14}\Ms$. Since in this case the uncertainties on the parameters, which 
are sensitive to the strong $w_0$-$w_a$ degeneracy, are much larger than in the previous case, the effect 
of the marginalization on $\fNL$ with a CMB prior is even smaller than in the case of time-independent $w$.
As before, however, marginalization over $\fNL$ with only the galaxy bispectrum prior substantially 
increases the error on $\sigma_8$.

As a final example, in Table~\ref{Table_2DE_nb10} we consider parameter constraints in the time-varying 
dark energy cosmology using the ten cluster mass bins. As noticed earlier in the case of the 4-parameter 
analysis, the $\sigma_8$-$\fNL$ degeneracy is significantly reduced and the expected constraints on
non-Gaussianity are still of the order of $\dfNL\sim 50$.

\section{Conclusions}
\label{secConclusions}

The success of the $\Lambda$CDM standard cosmological model in recent years has been nothing short of 
spectacular.  Upcoming surveys will either continue to confirm this model and constrain its parameters with 
unprecedented accuracy, or they will uncover discrepancies which will point the way toward improvements in 
our understanding of fundamental physics.  Two questions addressing cosmology beyond the standard model 
that have been the subject of substantial attention in recent years are: what is the nature of the dark 
energy which is driving the accelerated expansion of the universe? and second, are fluctuations in the 
primordial matter distribution Gaussian, and therefore consistent with the predictions of the simplest 
inflationary models? In this paper, we obtain a rough estimate of the success that one of the most 
promising cosmological probes, galaxy cluster counts, is likely to have in answering these fundamental 
questions.

We have assumed an ideal cluster survey with survey parameters expected for the upcoming SPT and DES 
projects. Our fiducial cosmological model includes both dark energy and primordial non-Gaussianity using 
popular parameterized models, $w$ and $w_a$ for the former, and $f_{NL}$ for the latter. Cluster number 
counts as a function of mass, redshift, and cosmology were estimated using a standard fit to simulations
\citep{Jenkins:2000bv}, which we adjusted to allow for mild non-Gaussian initial conditions, 
and all clusters above a threshold mass were considered to be "found" by our fiducial survey. We then 
performed a simple likelihood analysis on the cluster counts using priors from current WMAP and expected 
Planck and SDSS constraints on non-Gaussianity as well as approximate priors on the two other relevant 
cosmological parameters from other present and future data sets.

Our principal conclusion is that dark energy constraints are in all cases \emph{not} substantially 
degraded by primordial non-Gaussianity when the model parameterized by the constant $\fNL$ and current 
limits from CMB observations are assumed. This is true despite the fact that variations in $\fNL$ close
to current uncertainties induce differences in the mass function comparable in magnitude to variations 
of $10\%$ in the dark energy parameter $w$. A stronger degeneracy is observed instead between $\fNL$
and $\sigma_8$; in this case, the expected errors on $\sigma_8$ from future cluster surveys can be 
noticeably affected when non-Gaussianity is included in the analysis. 

\squeezetable
\begin{table}[t]
\caption{\label{Table_2DE_nb10} Same as Table~\ref{Table_2DE} but using 10 cluster mass bins.}
\begin{ruledtabular}
\begin{tabular}{l|cccc}
prior:            & $\dfNL=0$ & $\dfNL=5$ & $\dfNL=37$ & $\dfNL=145$                          \\
\hline\hline
\multicolumn{5}{l}{No priors on $\Omega_m$ and $\sigma_8$.}                                    \\
\hline
$\Delta w_0$      & $0.180$   & $0.180$~($0\%$)   & $0.186$~($3\%$)    & $0.197$~($9\%$)       \\
$\Delta w_a$      & $0.68$    & $0.68$~($0\%$)    & $0.71$~($4\%$)     & $0.78$~($15\%$)       \\
$\Delta \Omega_m$ & $0.0141$  & $0.0141$~($0\%$)  & $0.0151$~($7\%$)   & $0.0170$~($21\%$)     \\
$\Delta \sigma_8$ & $0.0078$  & $0.0079$~($1\%$)  & $0.0113$~($45\%$)  & $0.0158$~($100\%$)    \\
$\dfNL$           & -         & $5.0$             & $31$               & $51$                  \\
\hline\hline
\multicolumn{5}{l}{Gaussian priors: $\Omega_m = 0.27 \pm 0.035$, $\sigma_8 = 0.9 \pm 0.05$}    \\ 
\hline
$\Delta w_0$      & $0.167$   & $0.168$~($1\%$)   & $0.170$~($2\%$)    & $0.175$~($5\%$)       \\
$\Delta w_a$      & $0.64$    & $0.64$~($0\%$)    & $0.66$~($3\%$)     & $0.70$~($9\%$)        \\   
$\Delta \Omega_m$ & $0.0129$  & $0.0130$~($1\%$)  & $0.0136$~($5\%$)   & $0.0146$~($13\%$)     \\
$\Delta \sigma_8$ & $0.0072$  & $0.0073$~($1\%$)  & $0.0103$~($43\%$)  & $0.0138$~($92\%$)     \\
$\dfNL$           & -         & $5.0$             & $30$               & $48$                  \\
\hline\hline
\multicolumn{5}{l}{Gaussian priors: $\Omega_m = 0.27 \pm 0.0035$, $\sigma_8 = 0.9 \pm 0.01$}   \\ 
\hline
$\Delta w_0$      & $0.081$   & $0.081$~($0\%$)   & $0.83$~($2\%$)     & $0.84$~($4\%$)        \\
$\Delta w_a$      & $0.42$    & $0.42$~($0\%$)    & $0.42$~($0\%$)     & $0.42$~($0\%$)        \\   
$\Delta \Omega_m$ & $0.0033$  & $0.0033$~($0\%$)  & $0.0033$~($0\%$)   & $0.033$~($0\%$)       \\
$\Delta \sigma_8$ & $0.0021$  & $0.0023$~($10\%$) & $0.0049$~($130\%$) & $0.0063$~($200\%$)    \\
$\dfNL$           & -         & $5.0$             & $26$               & $35$                  \\
\hline\hline
\multicolumn{5}{l}{Fixed $\Omega_m=0.27$ and $\sigma_8=0.9$}                                   \\
\hline
$\Delta w_0$      & $0.067$   & $0.068$~($1\%$)   & $0.070$~($4\%$)   & $0.070$~($4\%$)        \\
$\Delta w_a$      & $0.37$    & $0.038$~($3\%$)   & $0.40$~($8\%$)    & $0.40$~($8\%$)         \\
$\dfNL$           & -         & $3.8$             & $5.9$             & $6.0$                  \\
\end{tabular}
\end{ruledtabular}
\end{table}

A secondary conclusion is that the cluster survey itself might have sufficient statistical power 
to provide a valuable cross check on any detection or non-detection of primordial non-Gaussianity in 
CMB experiments, particularly when information on the cluster distribution as a function of the mass
is taken into account. 

However, we must emphasize that we \emph{have not} attempted to include in our analysis any of the 
systematic and statistical errors in the clusters mass determination, which are likely to cause trouble 
for real surveys, as well as uncertainties on the predictions for the mass function, and our results must 
be interpreted with this in mind. We believe that our principal result should be quite robust, since any 
significant increase in the error budget will reduce constraining power on dark energy parameters and 
de-emphasize the confusion caused by any non-Gaussian initial conditions. On the other hand, the 
effectiveness of clusters as a cross check of primordial non-Gaussianity estimates from the CMB could be 
dramatically worsened, and should therefore be the subject of future work.

\acknowledgments

We thank Eiichiro Komatsu and Eric Linder for helpful comments on a earlier draft of the paper.
C.~V. would like to thank Martin White for useful discussions. This work was supported by the US 
Department of Energy at the University of Chicago and at Fermilab, by the Kavli Institute for Cosmological 
Physics at the University of Chicago and by NASA grant NAG5-10842 at Fermilab.

\newpage
\bibliography{ngclusters}

\end{document}